\documentclass{vgtc}                          

\ifpdf
  \pdfoutput=1\relax                   
  \usepackage{graphicx}                
  \DeclareGraphicsExtensions{.pdf,.png,.jpg,.jpeg} 
\else
  \usepackage{graphicx}                
  \DeclareGraphicsExtensions{.eps}     
\fi%

\graphicspath{{figures/}{pictures/}{images/}{./}} 
\usepackage{xcolor}
\usepackage{microtype}                 
\PassOptionsToPackage{warn}{textcomp}  
\usepackage{textcomp}                  
\usepackage{mathptmx}                  
\usepackage{times}                     
\usepackage{cite}                      
\usepackage{tabu}                      
\usepackage{booktabs}                  
\usepackage{paralist}
\newcommand{\systemname}{{GrieferLens}}

\onlineid{0}

\vgtccategory{Research}

\vgtcinsertpkg

\title{Towards an Exploratory Visual Analytics System for
Griefer Identification in MOBA Games}

\author{Zixin Chen\thanks{e-mail: zchendf@connect.ust.hk, co-first authors}\\ %
\scriptsize HKUST %
\and Shiyi Liu\thanks{e-mail: liushy@shanghaitech.edu.cn, co-first authors}\\ %
\scriptsize ShanghaiTech University %
\and Zhihua Jin\thanks{e-mail: zjinak@connect.ust.hk, co-first authors}\\ %
        \scriptsize HKUST %
\and Gaoping Huang\thanks{e-mail: gavinphuang@tencent.com, co-first authors}\\ %
     \scriptsize Tencent %
\and Yang Chao\thanks{e-mail: youngchao@tencent.com, co-corresponding authors}\\ %
\scriptsize Tencent %
\and Zhenchuan Yang\thanks{e-mail: zachyang@tencent.com}\\ %
\scriptsize Tencent %
\and Quan Li\thanks{e-mail: liquan@shanghaitech.edu.cn, co-corresponding authors}\\ %
\scriptsize ShanghaiTech University %
\and Huamin Qu\thanks{e-mail: huamin@cse.ust.hk}\\ %
\scriptsize HKUST %
}

\teaser{
  \centering
  \includegraphics[width=\linewidth]{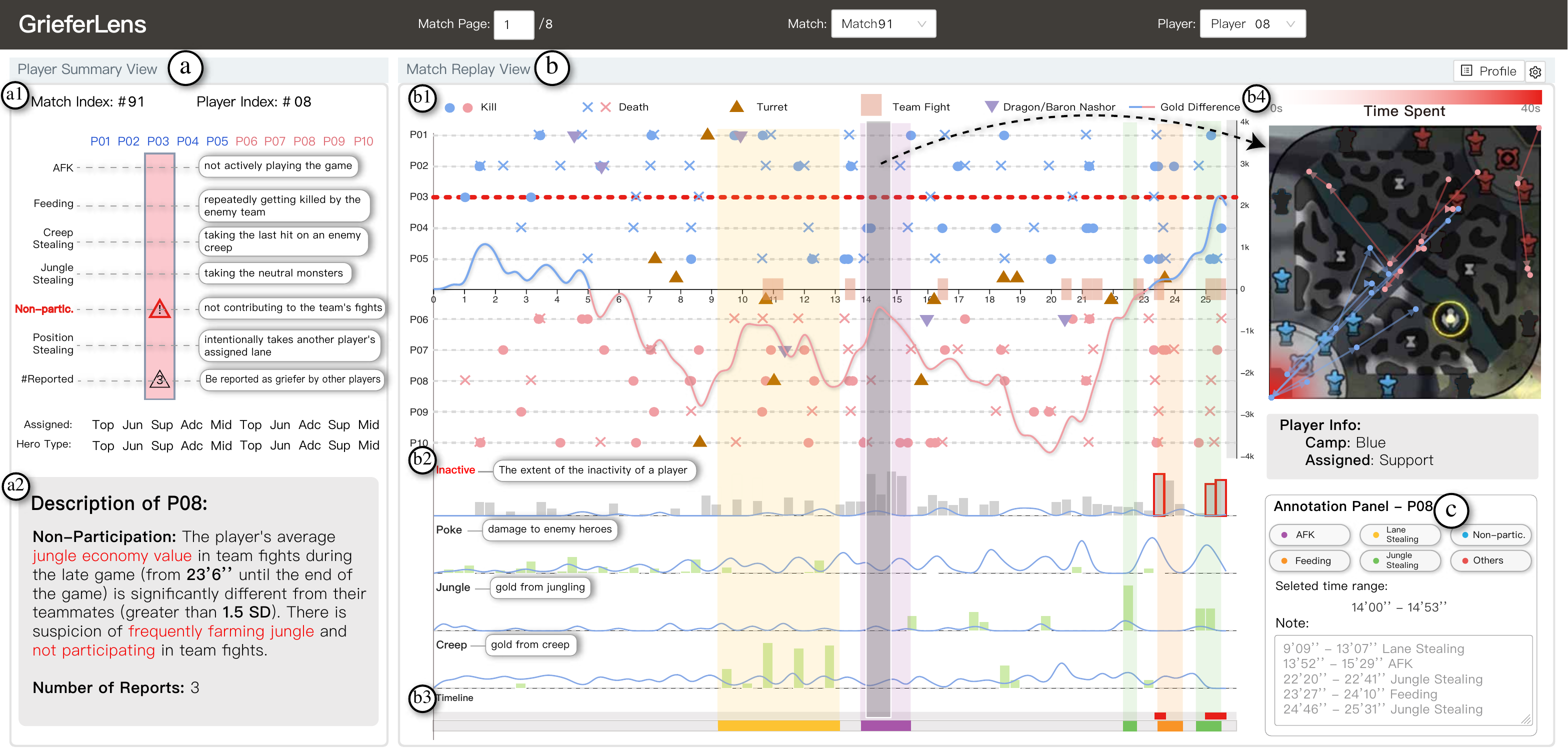}
  \caption{\textit{\systemname} is designed to help users identify and label griefers in MOBA games. It consists of Player Summary View and Match Replay View. (a) The Player Summary View provides an overview of the algorithmic analysis of each player's behavior. (b) Match Replay View summarizes the whole match progress and enables users to check the detailed events of one player and contextualize its trajectory with other players.}
  \label{fig:teaser}
}

\abstract{

Multiplayer Online Battle Arenas (MOBAs) have gained a significant player base worldwide, generating over two billion US dollars in annual game revenue. However, the presence of griefers, who deliberately irritate and harass other players within the game, can have a detrimental impact on players' experience, compromising game fairness and potentially leading to the emergence of gray industries. Unfortunately, the absence of a standardized criterion, and the lack of high-quality labeled and annotated data has made it challenging to detect the presence of griefers. Given the complexity of the multi-variant spatiotemporal data for MOBA games, game developers heavily rely on manual review of entire game video recordings to label and annotate griefers, which is a time-consuming process. To alleviate this issue, we have collaborated with a team of game specialists to develop an interactive visual analysis interface, called \textit{\systemname}. It overviews players' behavior analysis and synthesizes their key match events. By presenting multiple views of information, \textit{\systemname} can help the game design team efficiently recognize and label griefers in MOBA games and build up a foundation for creating a more enjoyable and fair gameplay environment. \\

} 

\CCScatlist{

  \CCScatTwelve{MOBA Games}{Griefers}{}{};
  \CCScatTwelve{Visualization}{Visual Analytics}{}{};
}

\begin{document}


\firstsection{Introduction}

\maketitle



\par Multiplayer Online Battle Arenas (MOBAs) are a popular genre of esports, with games like League of Legends (LoL) hosting professional tournaments on an international scale and generating over two billion US dollars in annual game revenue~\cite{Kokkinakis2020-uo}. However, some players engage in malicious behavior that disrupts the cooperative nature of the game, such as deliberately staying away from the keyboard (AFK) during gameplay. This behavior, known as griefing, can significantly impede the team's progress and disrupt the overall player experience, undermining the match's fairness and resulting in detrimental consequences for players~\cite{Kwak2015-ye}. 


\par Detecting griefers in MOBA games can be challenging due to the complex and multi-variant spatiotemporal game data and the absence of a standardized criterion for assessing player behavior, which necessitates a heavy reliance on expert judgment.
Previous studies have primarily focused on deciphering the reasons behind typical \textit{Snowballing} and \textit{Comeback} events, aiming to maintain a specific ratio that ensures fairness and engagement in the game \cite{7534855}. However, the field lacks comprehensive research into the detection of players who deliberately disrupt this balance. Currently, the game design team relies heavily on manual review of entire game video recordings to label and annotate, which is a time-consuming process requiring expertise. To address this problem, we developed a visual analytics system called \textit{GrieferLens} to aid users in identifying, labeling, and annotating griefers in MOBA games. The system integrates domain experts' knowledge to provide an overview analysis of players' behavior and synthesizes the key events during a match from both temporal and spatial perspectives. The trajectories of each player are also provided to help users understand the dynamics of match progress. Users can combine information from multiple perspectives to determine whether one player is a griefer and provide a label. The system's effectiveness and usability were evaluated via case studies.


\section{System Overview}
\textit{\systemname} focuses on identifying six types of griefers, as proposed and summarized by experts: AFK, feeding, lane stealing, jungle stealing, non-participation, and position stealing. The system contains two views and one annotation panel.
\par (a) The \textit{player summary view} displays whether each of the ten players in a game match is suspicious of exhibiting any of the six types of griefers (a1), which were determined using rule-based algorithms collaboratively designed and evaluated by our experts. Given the indeterministic nature of griefer behavior's definition, experts agreed that the first step in identifying griefers should be to use our algorithm's filtered output, which has been proven to reach a reasonable false-negative rate ($\leq 10\%$). Meanwhile, the number of times each player has been reported in this match, their default hero types and assigned positions, and a detailed description (a2) of their suspicion level based on the algorithm are also presented. 
\par (b) The \textit{match replay view} compresses an entire game into a single view by embedding the key events of all ten players (b1) and some quantitative metrics (b2) of the selected player into a timeline (b3). Suspicious time periods will be highlighted to guide users in identifying potential instances of griefers and finding evidence to support their labeling and annotating. The spatial data of players will be encoded in the map sub-view (b4). Users can easily review the game scene through the combination of the two sub-views. 
\par(c) Users can annotate problematic moments during the game using the \textit{annotation panel}, which supports different types of annotations like tags and text notes.

\section{Case Study}
A case is presented in Figure 1 to demonstrate how the experts explore and identify griefers in a game match.
\subsection{Key Event Labelling}
After selecting a match, the user first checks the \textit{player summary view} (a) to see the algorithm's assessments for each player. From the panel (a1), the user identifies a player \textit{P03} with suspicious behavior who didn't participate in some crucial team fights. The user discovers that \textit{P03}'s hero type is the ``Support" role, which indicates that its primary responsibility is to assist and protect teammates. The user then finds that \textit{P03} had a high jungle economy value compared to the team during the late stage of the game from the natural language explanation below. To investigate \textit{P03}'s behavior during the late stage, the user clicks on the two highlighted time periods (in green) in the \textit{match replay view} (b) and checks them. 
\par When checking the second highlighted time period, the user discovers that it shows the final team fight which led directly to the team's loss. While the rest of the team members were engaged in the fight against the opponents in the \textit{river area}, the suspicious player \textit{P03} was alone jungling in the \textit{enemy jungle area}. The red heatmap color shown in the map sub-view (b4) indicates that \textit{P03} stayed there for more than 30 seconds. As the ``Support" responsible for protecting teammates during the team fight, \textit{P03}'s behavior is a clear indication of malicious intent, which is highly likely to contribute to the game's failure. Consequently, the user concludes that \textit{P03} is indeed a griefer and labels it on the annotation panel. 
\subsection{Additional Annotation}
To further investigate and annotate the behavior of \textit{P03} throughout the whole match, the user returns to panel b2 to check for any additional malicious behavior. Using the highlighted ``Inactive'' metric, which quantifies a player's positive contribution to the team every 20 seconds, the user easily identifies that \textit{P03}'s contribution to the team was especially low from 13 to 14 minutes. After selecting this time period, the user examines the map sub-view (b4) and identifies that \textit{P03} kept moving between the \textit{Towers} and the \textit{Crystal}, while also staying in the \textit{Fountain} for more than 20 seconds without any apparent reason. Consequently, the user annotates this period of time and adds comments in the annotation panel (c), completing the investigation of the griefer's behavior in this match.

\section{Expert Feedback}

Following the development of our system, we sought feedback from several experts who compared its efficiency and accuracy to their traditional approach. The results showed that using our system \textit{\systemname}, experts were able to annotate in approximately 5 minutes, a significant improvement over the 25 minutes required to watch a full game video (baseline). Additionally, the annotation results were highly consistent, further demonstrating the effectiveness of our system. The most troublesome part of watching videos was identifying key moments in the video, while experts claimed that ``\textit{the suspicious time range and the timeline summarizing the key events were very helpful.}'' Additionally, the bar charts of ``Inactive"  helped them ``\textit{easily identify outliers at a glance}''. They even discovered some previously mislabeled annotations in the process.

\section{Conclusion and Future Work}
To mitigate the burden of the labelling and annotation of various types of griefers in MOBA games, we closely collaborated with experts to identify six types of griefers and iteratively designed \textit{GrieferLens}. \textit{GrieferLens} significantly improves the efficiency of these tasks by providing suspicious event prompts and visual summaries of spatiotemporal game data. Feedback from experts suggests that the system is effective, and future work will involve conducting larger user studies as evaluations and exploring extra valuable system functionalities. Overall, \textit{GrieferLens} represents a promising approach to addressing the issue of griefer detection in MOBA games and has the potential to significantly improve the overall gameplay experience for all players.


\bibliographystyle{abbrv-doi}

\bibliography{template}

\begin{thebibliography}{1}

\bibitem{Kokkinakis2020-uo}
A.~V. Kokkinakis, S.~Demediuk, I.~Nölle, O.~Olarewaju, S.~Patra, J.~Robertson,
  P.~York, A.~P. Pedrassoli~Chitayat, A.~Coates, D.~Slawson, P.~Hughes,
  N.~Hardie, B.~Kirman, J.~Hook, A.~Drachen, M.~F. Ursu, and F.~Block.
\newblock {DAX}: Data-driven audience experiences in esports.
\newblock In {\em Proceedings of the {ACM} International Conference on
  Interactive Media Experiences}, pp. 94--105, 2020. doi: {{%
10\hspace{.1pt}\discretionary{.}{%
}{.}\hspace{.4pt}1145\discretionary{/}{%
}{/}3391614\hspace{.1pt}\discretionary{.}{%
}{.}\hspace{.4pt}3393659}}


\bibitem{Kwak2015-ye}
H.~Kwak, J.~Blackburn, and S.~Han.
\newblock Exploring cyberbullying and other toxic behavior in team competition
  online games.
\newblock In {\em Proceedings of the 33rd Annual {ACM} Conference on Human
  Factors in Computing Systems}, pp. 3739--3748, 2015. doi: {{%
10\hspace{.1pt}\discretionary{.}{%
}{.}\hspace{.4pt}1145\discretionary{/}{%
}{/}2702123\hspace{.1pt}\discretionary{.}{%
}{.}\hspace{.4pt}2702529}}


\bibitem{7534855}
Q.~Li, P.~Xu, Y.~Y. Chan, Y.~Wang, Z.~Wang, H.~Qu, and X.~Ma.
\newblock A visual analytics approach for understanding reasons behind
  snowballing and comeback in moba games.
\newblock {\em IEEE Transactions on Visualization and Computer Graphics},
  23(1):211--220, 2017. doi: {{%
10\hspace{.1pt}\discretionary{.}{%
}{.}\hspace{.4pt}1109\discretionary{/}{%
}{/}TVCG\hspace{.1pt}\discretionary{.}{%
}{.}\hspace{.4pt}2016\hspace{.1pt}\discretionary{.}{%
}{.}\hspace{.4pt}2598415}}


\end{thebibliography}
\end{document}